\documentclass[prb,twocolumn,superscriptaddress,showpacs]{revtex4}
\usepackage{mathrsfs}
\usepackage{amsmath,amsfonts,amssymb}
\usepackage{epsfig}
\usepackage{graphicx}
\usepackage{wasysym}
\usepackage{bbm}
\usepackage{psfrag}
\usepackage{color}
\usepackage{pstool}
\usepackage{braket}
\usepackage[dvips, bookmarks, colorlinks=true, plainpages = false, citecolor = blue, linkcolor = blue, urlcolor = blue, filecolor = blue]{hyperref}
\newcommand \be{\begin{equation}}
\newcommand \ee{\end{equation}}
\newcommand \bes{\begin{equation*}} 
\newcommand \ees{\end{equation*}}
\newcommand \bea{\begin{eqnarray}}
\newcommand \eea{\end{eqnarray}}
\newcommand \bsea{\begin{subequations}\begin{eqnarray}} 
\newcommand \esea{\end{eqnarray}\end{subequations}}
\newcommand \beas{\begin{eqnarray*}} 
\newcommand \eeas{\end{eqnarray*}}
\newcommand \bfg{\begin{figure}}
\newcommand \efg{\end{figure}}
\newcommand \bfgs{\begin{figure*}} 
\newcommand \efgs{\end{figure*}}
\newcommand \bwt{\begin{widetext}}
\newcommand \ewt{\end{widetext}}

\def\pmat#1{\left(\begin{matrix}#1\end{matrix}\right)}

\begin{document}
\title{Emergence of photo-induced multiple topological phases on square-octagon lattice }
\author{Arghya Sil}\email{arghyasil36@gmail.com}
\affiliation{Department of Physics, Jadavpur University, 188 Raja Subodh Chandra Mallik Road, Kolkata 700032, India}
\author{Asim Kumar Ghosh}\email{asimkumar96@yahoo.com}
\affiliation{Department of Physics, Jadavpur University, 188 Raja Subodh Chandra Mallik Road, Kolkata 700032, India}
\begin{abstract}
In this study, a tight-binding model on square-octagon lattice 
with nearest-neighbour and next-nearest-neighbour 
hoppings  is considered. The system is topologically trivial although 
it exhibits quadratic band-touching points in its band-structure. 
It is shown that the system drives towards topologically non-trivial phases 
as soon as it is exposed to monochromatic circularly 
polarized light. Multiple topological phases are found to 
emerge with the variation of amplitude of light. An effective time-independent Hamiltonian 
of the system is obtained following the Floquet-Bloch formulation.  
Quasi-energy band-structures along with definite values of 
Chern numbers for respective bands are obtained. 
Characterization of topological 
phases are made in terms of band structure and Chern numbers. In addition, 
Hall conductance and topologically protected chiral 
edge states are obtained to explain the 
topological properties. 
Density of states is obtained in every case. 
The system undergoes a series of topological phase transitions 
among various topological phases upon variations of both 
amplitude of light and hopping strengths. It exhibits Chern insulating and 
Chern semimetallic phases depending on the values of lattice filling. 

\end{abstract}
\maketitle
\date{today}

\section{Introduction}
Topological phases of matter have become the centre of attraction among the 
researchers in recent years \cite{Kane2}. A considerable portion of 
these investigations includes the 
study of topological insulators (TI) or quantum spin Hall (QSH) phase, 
Chern insulators or quantum anomalous Hall phase, topological Dirac or Weyl
semimetals, to name a few. Depending on the nature of time reversal symmetry (TRS), 
particle-hole symmetry and chiral symmetry of the systems, 
ten classes of topological insulators have been proposed \cite{Ryu}. 
A transition from either trivial band-insulators or semimetals to non-trivial 
topological phases traditionally occurs in two different ways. 
Either it is driven by artificial gauge-field in case of Chern insulators, 
where each energy band is characterized by the integral Chern number ($C$) 
\cite{TKNN}, or spin-orbit coupling (SOC) in case of QSH effect, 
where the resulting system is characterized by a non-trivial
$Z_2$ invariant \cite{Kane2}.
Haldane first demonstrated the emergence of Chern insulating (CI) phase  
in a two-band model 
on graphene by introducing an artificial phase coupled with next-nearest 
neighbour hopping, which breaks TRS \cite{Haldane}. 
On the other hand, QSH was also initially proposed on graphene,  
but with intrinsic SOC instead, providing a net TRS \cite{Kane1}. 

Besides those two ways, another route for driving the 
system into non-trivial topological phases has been established 
by a series of recent investigations. 
That particular state of matter is known as Floquet 
topological insulator (FTI) which is obtained by perturbing the 
system with a coherent periodic photo-irradiations \cite{Oka,Inoue1}. 
Topological classification of periodically driven quantum systems has been 
formulated \cite{Kitagawa}. 
So, in order to study the FTI phase in a tight-binding system, one has to develop an 
appropriate Floquet-Bloch theory on it. 
A section of investigations is now devoted to find  
the existence of FTI phase 
within the systems those were otherwise declared as topologically 
trivial. 
These kind of investigations may help the experimentalists
to develop suitable techniques for probing the existence of 
novel topological phases within the materials. 
In addition, associated topological phase 
transitions can also be demonstrated by modulating the appropriate 
parameters of the irradiation \cite{Inoue2}.
    
Emergence of quasi-continuous edge-states on the surface 
of the material is one of the most significant evidence of any 
kind of TI. These topologically protected edge states 
are obtained for finite system in open boundary condition (OBC) 
and in general they tend to obey the `bulk-boundary correspondence' 
rule \cite{Hatsugai1,Hatsugai2}. Following the successful achievement 
in the field of conventional electronic TI's, their bosonic analogs,
namely topological magnon insulators have also drawn attention  
in more recent years \cite{Mook,Owerre3,Kim}.

Existence of FTI phase was also predicted in graphene for the first time. 
Since then there have been several studies on Floquet topological
phase on honeycomb lattice \cite{Mitra1,Kundu,Rigol,Torres,Li} in addition to    
other two-band systems \cite{Saha,Galitski,Ezawa,Platero}.
It has been shown that the effect of periodically driven circularly 
polarized light leads to a non-trivial mass term similar to that found in  
Haldane model \cite{Moessner}. This mass term now becomes a function of
both the intensity and the state of polarization of light. 
So, detection of a particular non-trivial topological phase requires  
appropriate value of intensity and definite state of polarization of the light. 
In addition to the honeycomb lattice, 
effect of photo-irradiation 
is studied before on three-band tight-binding models formulated on 
kagom\'e \cite{Fiete1} and Lieb \cite{Ren} lattices. 
Again, in these tight-binding models, photo-irradiation 
opens up gaps in the otherwise gapless spectrum 
and the resulting energy-bands are found to emerge simultaneously with non-zero Chern 
numbers. 
Interestingly, in two-dimensional Floquet systems, 
evidence of chiral edge states is found even though 
the Chern numbers of the respective bulk bands are zero, 
which indicates the violation of `bulk-boundary correspondence' 
rule for that particular system \cite{Kitagawa,Rudner}. 
Therefore, determination of both $C$ for the bulk-bands and edge states 
between the respective bands are necessary to confirm the topological phases. 
Recently, the bosonic counterpart of 
FTI phase has been formulated in magnetic systems \cite{Owerre1}. 

In this work, we pay attention to the four-band tight-binding model 
on square-octagon lattice. 
It has been reported that this model  
exhibits $Z_2$ band-insulating and CI phases  
in the presence of spin-orbit coupling \cite{Fiete2} 
and external magnetic flux \cite{Pal,Gong}, respectively. 
Topological phase transition (TPT) driven by 
SOC and exchange field are 
also found on this lattice \cite{Yang}. 
Outcome of those studies based on square-octagon lattice 
along with the emergence of photoinduced topological phases on 
kagom\'e and Lieb lattices 
motivate us to search the FTI phase on this particular lattice. 

The intrinsic tight-binding model on square-octagon lattice 
has an interesting band-structure which includes 
a pair of flat band and five different band-touching points. 
The coherent photo-irradiation 
perturbs the band structure in such a manner that either true or pseudo 
band-gaps appear depending on the value of its amplitude. 
System undergoes a series of
topological phase transition with the variation of amplitude, 
where successive band-touching and gap-opening processes occur 
at the transition points separating the different topological phases. 
As a result, multiple topological phases are found to emerge in the 
parameter space, which can be categorized as either Chern semimetallic (CSM) or CI phases 
depending on the nature of band-gap. CSM (CI) phase appears in 
case of pseudo (true) band-gap. 
Rapid variation of topological phases with respect to the amplitude of 
irradiation is observed which is not reported before for the cases of 
honeycomb, kagom\'e and Lieb lattices. 
Chiral edge states are found to appear in finite 
system according to the `bulk-boundary correspondence' rule which  
confirms the existence of non-trivial topological phases. 

The plan of the paper is as follows. In section \ref{model}, 
square-octagon lattice is described and the tight-binding Hamiltonian
is formulated. Outline of Floquet theory and development of 
Floquet-Bloch effective Hamiltonian
are presented in section \ref{Floquet}. 
Detailed characterization of various topological phases and 
transition among them along with explanations are provided in 
section \ref{Properties}. Section \ref{discussions} contains discussion
on the theme of this work.
   
\section{Model Hamiltonian and Band Structure}
\label{model}
\begin{figure}[t]
 \centering
 \psfrag{j}{\text {\tiny {\bf$ \delta_5 $}}}
 \psfrag{b}{\text {\tiny {\bf$ \delta_6 $}}}
 \psfrag{c}{\text {\tiny {\bf$ \delta_1 $}}}
 \psfrag{h}{\text {\tiny {\bf$\delta_2 $}}}
 \psfrag{d}{\text {\tiny{$\Gamma$}}}
 \psfrag{e}{\text {\tiny{K}}}
 \psfrag{f}{\text {\tiny{M}}}
 \psfrag{x}{\text {\scriptsize{$ x $}}}
 \psfrag{y}{\text {\scriptsize{$ y $}}}
  \psfrag{E}{\text {\tiny {\bf $\delta_3$}}}
 \psfrag{F}{\text {\tiny {\bf $ \delta_4$}}}
 \psfrag{(a)}{\text{{(a)}}}
 \psfrag{(b)}{\text{\scriptsize{$(b)$}}}
\includegraphics[width=6.2cm,height=6.2cm]{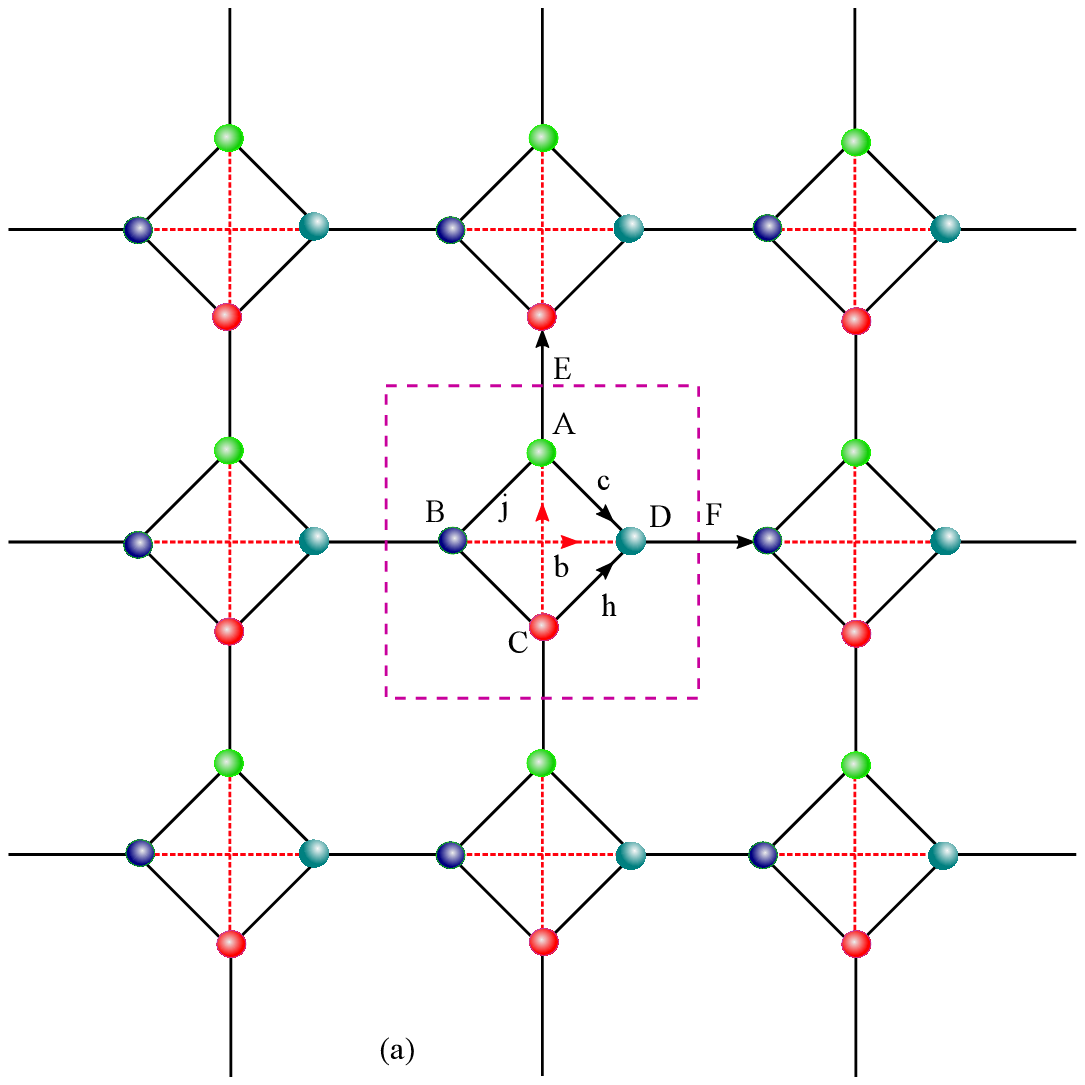} 
\includegraphics[width=1.8cm]{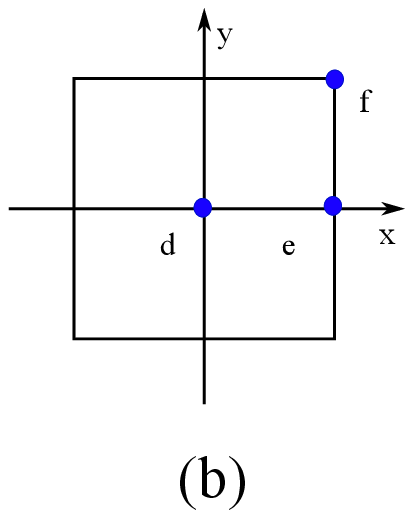} 
\caption{(Color online) (a) The geometry of square-octagon lattice. $A,B,C,D$ denote four 
different sites in the unit cell which is shown by large square 
(pink dashed lines).
Strength of NN hopping along sides (black solid lines) 
of small square is $t_1$.  
Strength of NNN hopping along diagonals (red dotted lines) 
of small square is $t_2$. 
$\boldsymbol \delta_1, \boldsymbol \delta_2,\boldsymbol \delta_3, \boldsymbol \delta_4$ 
are the NN vectors and 
$\boldsymbol \delta_5, \boldsymbol \delta_6$ are the NNN vectors. 
(b) The first Brillouin zone of the lattice.  High-symmetry points 
$\Gamma$, K and M have been marked.}
\label{Lattice}
\end{figure}

We consider a two-dimensional square-octagon lattice model 
consisting of two elementary plaquettes: square and 
octagon as shown in Fig \ref{Lattice}. 
The same lattice is noted before as 1/5-depleted square lattice \cite{Ueda} and 
Archimedean lattice T11 \cite{Yu} in the context of investigations of 
several other properties based on it. 
The coordination number of this non-Bravais lattice 
is three which is similar to that of honeycomb lattice, another 
non-Bravais one. However, this lattice can be considered as 
composed of four interpenetrating square sub-lattices each of which is 
formed by lattice points those are drawn by four different colours in Fig 
\ref{Lattice}.  
The tight binding Hamiltonian considering nearest-neighbour (NN) 
and next-nearest-neighbour (NNN) interactions can be written as 
\bea
 H\!\!&=&\!\!\! -\!\!\!\!\sum_{(m,n)}\!\!\bigg[t_1\!\sum_{\langle ij\rangle} c_{m,n,i}^{\dagger}\,
c_{m,n,j}\! +\!t_2\!\sum_{\langle\langle ij\rangle\rangle}
  c_{m,n,i}^{\dagger}\,c_{m,n,j}\!+\!H.c \bigg]. \nonumber\\
\eea
The first summation $(m,n)$ runs over the unit cell indices, while 
the second summations $\langle \cdot\rangle$ and 
$\langle\langle \cdot\rangle\rangle$ run over NN and NNN pairs, respectively. 
$c_{m,n,i}^{\dagger} \,(c_{m,n,i})$ is the creation (annihilation) operator
 for an electron at the $i$-th site of the $(m,n)$-th unit cell. 
 $t_1$ is the strength of NN hopping while $t_2$ is that of 
NNN hopping those occur along the diagonals of each square plaquette. 
By setting the distance between NN points to be unity, 
four NN vectors are  defined as 
$\boldsymbol \delta_1 = 1/\sqrt{2}\left(1,-1\right) ,\;
 \boldsymbol \delta_2 = 1/\sqrt{2}\left(1,1\right) ,\; \boldsymbol \delta_3 = \left(0,1\right)$, and $\boldsymbol \delta_4 = \left(1,0\right)$. 
Similarly, two NNN vectors are $\boldsymbol \delta_5 = 
\sqrt{2}\left(0,1\right)$, and
$\boldsymbol \delta_6 = \sqrt{2}\left(1,0\right)$.  
 The lattice translation vectors are $\boldsymbol a_1 = \left(1+\sqrt{2},0\right) $ and $\boldsymbol a_2 = \left(0,1+\sqrt{2}\right)$. 
 Position of each unit cell is defined by $\boldsymbol R\left(m,n\right) = m\boldsymbol a_1+n\boldsymbol a_2$.
 We set $t_1=1.0$ throughout the article. 
 
 Fourier transforming to the momentum space, the Hamiltonian becomes 
 $H = -\sum_{\textbf{k}} \psi_{\textbf{k}}^{\dagger}\, H(\textbf{k})\, \psi_{\textbf{k}} $, where 
   \begin{equation}
 \begin{aligned}
  H(\textbf{k})\! =\! \pmat{0&t_1&t_1 e^{ik_2}\!+\!t_2&t_1 \\ t_1&0&t_1&t_1e^{-ik_1}\!+\!t_2 \\ t_1 e^{-ik_2}\!+\!t_2 &t_1 &0&t_1 \\
         t_1&t_1e^{ik_1}\!+\!t_2&t_1&0}
 \end{aligned}
\end{equation}
with $\psi_{\textbf{k}}= {\left(A_{\textbf{k}},B_{\textbf{k}},C_{\textbf{k}},D_{\textbf{k}}\right)}^{T} $,  
where $\alpha_{\textbf{k}}$ with $\alpha = A,B,C,D $ are the electron 
annihilation operators on the four basis sites in the square unit cell, 
$ k_1=\boldsymbol k\cdot \boldsymbol a_1$ and $ k_2=\boldsymbol k\cdot \boldsymbol a_2$.
\begin{figure}[t]
\centering
\psfrag{E}{\text{\scriptsize{Energy}}}
\psfrag{K1}{$k_1$}
\psfrag{K2}{$k_2$}
\psfrag{$3$}{$\pi$}
\psfrag{c0}{\text{\tiny{\bf  {$C=0$}}}}
\psfrag{c2}{\text{\tiny{\bf  {$C=2$}}}}
\psfrag{c-2}{\text{\tiny{\bf  {$C=-2$}}}}
\psfrag{c1}{\text{\tiny{\bf  {$C=1$}}}}
\psfrag{c-1}{\text{\tiny{\bf  {$C=-1$}}}}
\psfrag{p}{\text{\tiny{$E_1$}}}
\psfrag{q}{\text{\tiny{$E_2$}}}
\psfrag{r}{\text{\tiny{$E_3$}}}
\psfrag{s}{\text{\tiny{$E_4$}}}
\includegraphics[width=4cm,height=5cm,trim={2.0cm .7cm 3.2cm .5cm}]{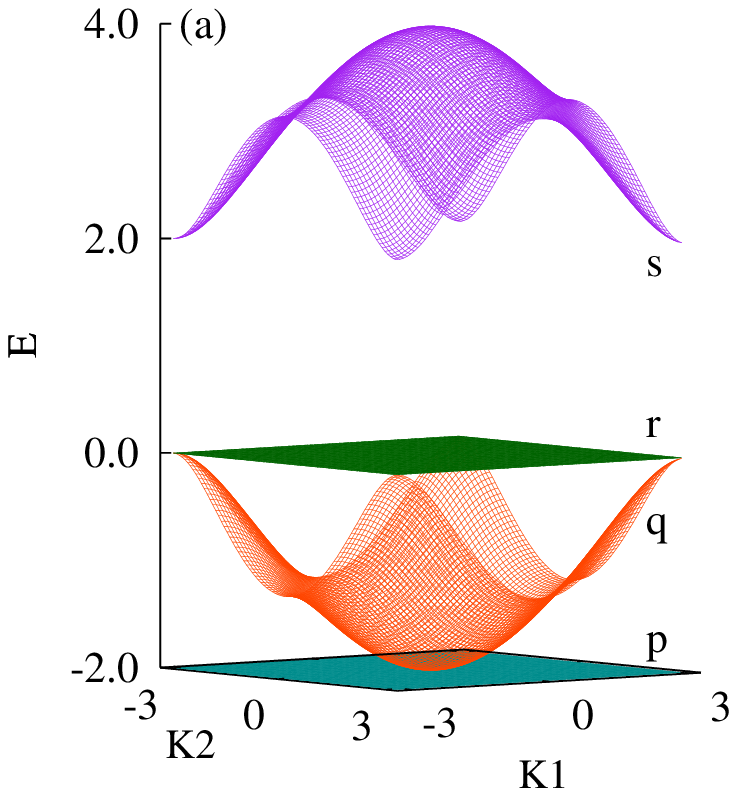} 
\includegraphics[width=4cm,height=5cm,trim={2.0cm .5cm 3.2cm .5cm}]{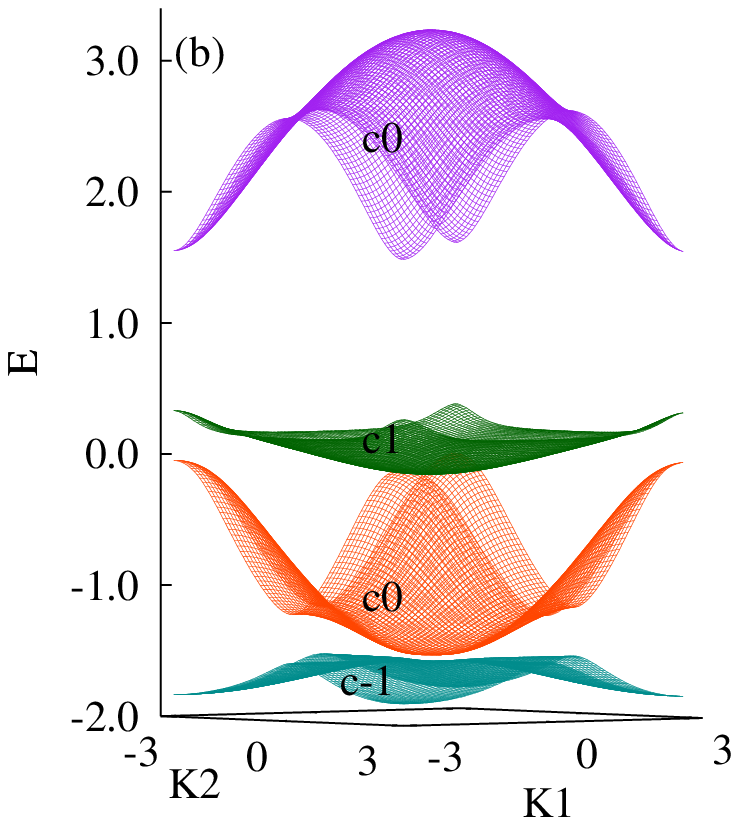}  %
\vskip .5 cm
\includegraphics[width=4cm,height=5cm,trim={2.0cm .5cm 3.2cm .5cm}]{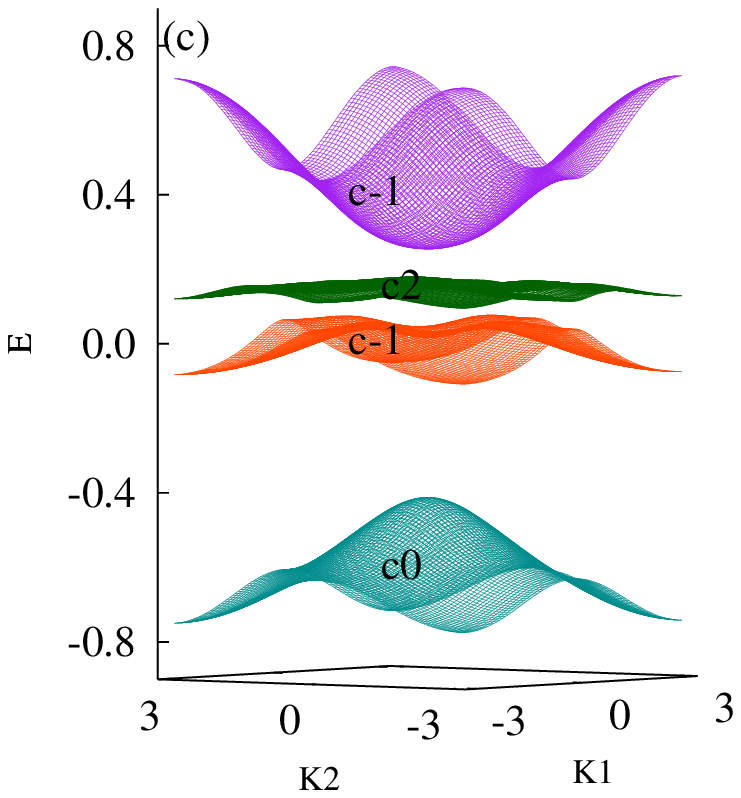}
\includegraphics[width=4cm,height=5cm,trim={2.0cm .7cm 3.5cm .5cm}]{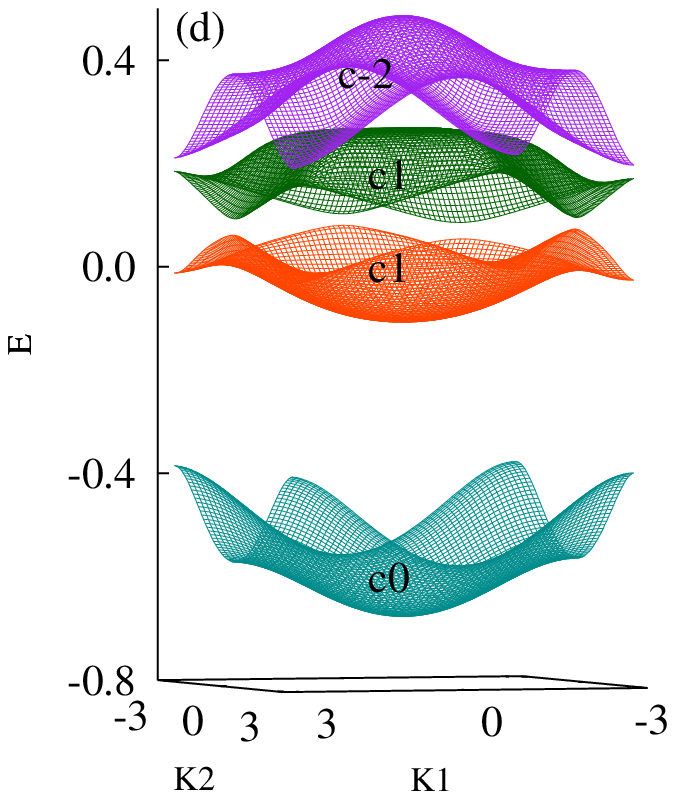} 
\caption{(Color online) (a) Band diagram of tight-binding square-octagon lattice without any
external perturbation when $t_2=1.0$. Floquet-Bloch band diagram in the 
presence of circularly polarized light for $t_2=1.0$  
$\Omega=7, A_0=0.8$ (b), $t_2=1.0, \Omega=10, A_0=2.1$ (c), $t_2=0.6, \Omega=10, A_0=2.7$ (d).}
\label{band3d}
\end{figure}

Diagonilizing the Hamiltonian we get two flat bands, ($E_1,E_3$) 
and two dispersive bands with energies 
$E_{2,4}(\textbf{k}) = 1 \pm \sqrt{5+2\cos{k_1}+2\cos{k_2}} $, 
as shown in Fig \ref{band3d} (a), when $t_2=1.0$. 
The bands are denoted as $E_1,E_2,E_3$ and $E_4$ in the 
ascending order of energy.
We note that, the lower two bands touch at the $\Gamma$
point while the energy bands $E_2(\textbf{k})$ and $E_3$ touch at 
four M points of the first Brillouin zone (BZ). All these are 
quadratic band touching points since the energy of the lower dispersive 
band, $E_2(\textbf{k})$, is found to become proportional to the square of
the wave-vectors when expanded in the vicinity of those points as shown below. 
\begin{equation}
 \begin{aligned}
 {E_{2}(k_1,k_2)|}_{k_1,k_2 \to 0} &= -2 + \frac{{k_1}^{2}+{k_2}^{2}}{6},\\
  {E_{2}(k_1,k_2)|}_{k_1,k_2 \to \pi} &= -\frac{{(k_1-\pi)}^{2}}{2}-\frac{{(k_2-\pi)}^{2}}{2}. 
 \end{aligned}
\end{equation}

The model without NNN hopping has been studied before 
in the absence of circularly polarized light. 
In this case, Dirac cones 
are found to appear at $\Gamma$ and M points when the intra- 
and inter-square-plaquette NN hopping strengths are different. \cite{Ueda}. 
\begin{figure*}[t]
\centering
\psfrag{DOS}{\text{\tiny{DOS}}}
\psfrag{E}{\text{\scriptsize{Energy}}}
\psfrag{G}{$\Gamma$}
\psfrag{c0}{\text{\tiny{\bf  {$C=0$}}}}
\psfrag{c2}{\text{\tiny{\bf  {$C=2$}}}}
\psfrag{c-2}{\text{\tiny{\bf  {$C=-2$}}}}
\psfrag{c1}{\text{\tiny{\bf  {$C=1$}}}}
\psfrag{c-1}{\text{\tiny{\bf  {$C=-1$}}}}
\includegraphics[width=15cm]{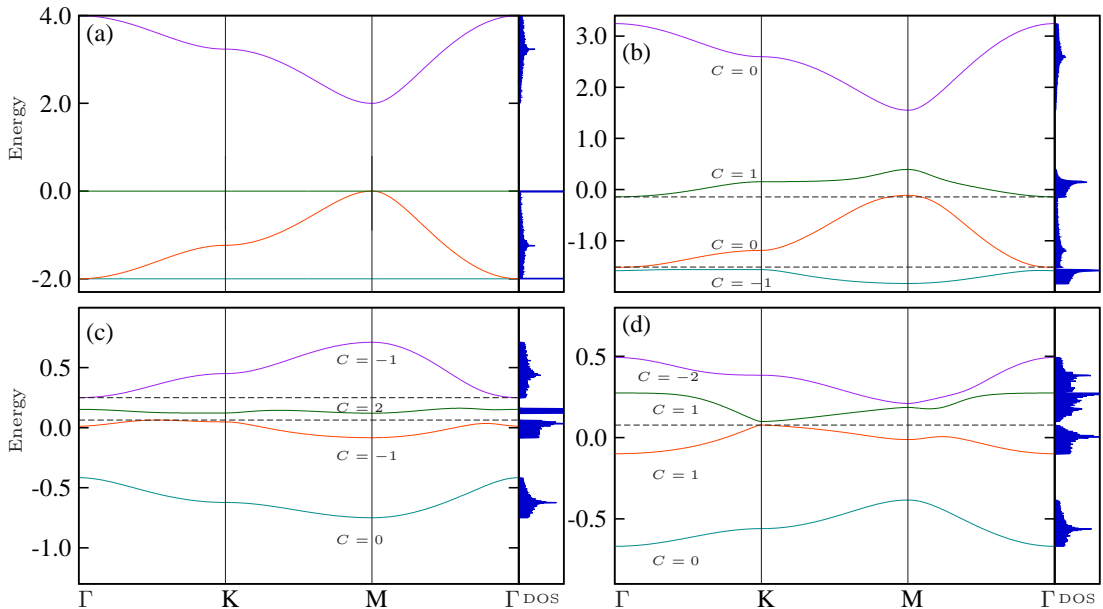}  %
\caption{(Color online) Floquet-Bloch band diagram of the tight-binding square-octagon lattice 
along the high-symmetry points of BZ.
(a) Without any external perturbation for $t_2=1.0$.     
(b) In the presence of circularly polarized light
when $t_2=1.0, \Omega=7, A_0=0.8$. System clearly exhibits the CI phase for 1/4 filling
and CSM phase for 1/2 filling as 
true band gap only exists between the two lower bands.
(c) When $t_2=1.0, \Omega=10, A_0=2.1$, the system exhibits the CI 
phase for both 1/2 and 3/4 fillings. 
True band gap exists between all the bands. 
(d) When $t_2=0.6, \Omega=10, A_0=2.7$, CI and CSM 
phases appear for 1/2 and 3/4 fillings, respectively.
Chern numbers of the respective energy-bands are noted.
Density of states are also shown in the right panels.
Horizontal dashed lines are drawn for the determination 
of true and pseudo band-gaps.}
\label{band2d}
\end{figure*}
\section{Derivation of Effective Hamiltonian by Floquet-Bloch Theory}
\label{Floquet}
In order to apply the Floquet-Bloch theory, the system is exposed 
to periodically driven monochromatic laser light which 
couples the momentum with the vector potential of the 
incident field $\textbf{A}\!=\!\textbf A_0\left[\sin(\Omega t),\cos(\Omega t) \right] $ through the substitution $\textbf {k}\rightarrow \textbf {k} 
+\textbf  A(t)$. $\textbf A_0$ and $\Omega$ are respectively amplitude and 
frequency of the laser light. 
Values of the relevant constants are assumed to be unity, so, 
$ \hbar =1,e=1,c=1 $. Therefore, the time-dependent Hamiltonian looks like,  
\begin{equation}
 \begin{aligned}
H(\textbf{k},t) &= -t_1 \sum_{\textbf{k}} \left[ {A_{\textbf{k}}}^{\dagger} B_{\textbf{k}} e^{-i\textbf A(t) \cdot \mathbf{\delta_2}}
+ {A_{\textbf{k}}}^{\dagger} D_{\textbf{k}} e^{i\textbf A(t) \cdot \mathbf{\delta_1}} \right. \\
&+\left. {B_{\textbf{k}}}^{\dagger} D_{\textbf{k}} e^{-i \textbf k \cdot \textbf
 R \left(1,0\right)}e^{-i\textbf A(t) \cdot \mathbf{\delta_4}}  
+ {B_{\textbf{k}}}^{\dagger} C_{\textbf{k}} e^{i\textbf A(t) \cdot \mathbf{\delta_1}} \right. \\
&+ \left. {A_{\textbf{k}}}^{\dagger} C_{\textbf{k}} e^{i \textbf k \cdot \textbf R 
\left(0,1\right)}e^{i\textbf A(t) \cdot \mathbf{\delta_3}}
+{C_{\textbf{k}}}^{\dagger} D_{\textbf{k}} e^{i\textbf A(t) \cdot \mathbf{\delta_2}}  \right] \\
&-t_2 \sum_{\textbf{k}} \left[  {A_{\textbf{k}}}^{\dagger} C_{\textbf{k}} e^{-i\textbf A(t) \cdot \mathbf{\delta_5}} 
+   {B_{\textbf{k}}}^{\dagger} D_{\textbf{k}} e^{i\textbf A(t) \cdot \mathbf{\delta_6}} \right].
\end{aligned}
\label{Ht}
\end{equation}
Now, the Hamiltonian becomes periodic with time period $ T=2\pi/\Omega $ like 
$ H(\textbf{k},t) = H(\textbf{k},t+T)$ because of the periodicity of 
${ \textbf A}(t) $ and monochromaticity of the laser 
light. This time-periodicity allows one to map the time-dependent 
problem into an effective time-independent
problem by virtue of Floquet theory \cite{Floquet} as described below.
By performing the Fourier transform in the time space, the periodic 
Hamiltonian can be expressed as 
$ H(\textbf{k},t) = \sum_{p=-\infty}^{\infty} e^{ip\Omega t}H_{p}
\left(\textbf{k}\right) $, 
where $ H_{p}\left(\textbf{k}\right) = \frac{1}{T}\int_{0}^{T} 
e^{-ip\Omega t}H(\textbf{k},t)\,dt =H_{-p}^{\dagger}\left(\textbf k\right)$
is the $p$-th order Fourier component.
By considering the time-dependent Schr\"odinger equation for the system, 
\begin{equation}
\begin{aligned}
 i\hbar\, \frac{d}{dt}\ket{\psi (\textbf{k},t)}=H(\textbf{k},t)\,\psi(\textbf{k},t),
 \label{2}
 \end{aligned}
\end{equation}
one may have the solution for the $\alpha$-th band as 
 \begin{equation}
\begin{aligned}
 \psi_{\alpha}(\textbf{k},t)=e^{i\epsilon_{\alpha}(\textbf{k})t} 
\phi_{\alpha}(\textbf{k},t), 
 \label{3}
 \end{aligned}
\end{equation}
where  $\phi_{\alpha}(\textbf{k},t)$ and $\epsilon_{\alpha}(\textbf{k})$ 
are known as the time-periodic Floquet-Bloch wave function and 
Floquet quasi-energy of that band, respectively. 
Floquet-Bloch wave function can be expressed in the Fourier space as 
$  \phi_{\alpha}(\textbf{k},t) = \sum_{p=-\infty}^{\infty} 
e^{ip\Omega t}\phi_{\alpha,p}\left(\textbf{k}\right) $. 
Temporal periodicity of the Hamiltonian  
leads to the appearance of periodicity of the quasi-energies  
in the frequency space with period $\Omega$, which, on the other hand,
gives rise to the existence of 
quasi-energy BZ bounded within the limits $-\Omega/2<\epsilon<\Omega/2$. 
This result is similar to the appearance of periodicity for 
the energy of Bloch electron in the momentum space. 

By using Eqs (\ref{2}, \ref{3}) and defining 
the Floquet operator as $ H_{F}(\textbf{k},t) = 
H(\textbf{k},t)-i\partial_{t} $, the following eigenvalue 
equation for the Floquet-Bloch wave function is obtained.  
\begin{equation}
\begin{aligned}
 H_{F}(\textbf{k},t) \ket{\phi_{\alpha}(\textbf{k},t)} = \epsilon_{\alpha}(\textbf{k})\ket{\phi_{\alpha}(\textbf{k},t)}.
 \label{4}
 \end{aligned}
\end{equation}
Solution of this equation in the Fourier space leads to a time-independent 
Floquet eigenvalue problem,  
\begin{equation}
\begin{aligned}
 \sum_q \left[H_{p-q}(\textbf{k}) + q\,\Omega \,\delta_{p,q}\right]
\ket{\phi_{\alpha,q}(\textbf{k})}
  = \epsilon_{\alpha}(\textbf{k})\ket{\phi_{\alpha,p}(\textbf{k})}.
 \label{5}
 \end{aligned}
\end{equation}
Floquet theory finally leads to a set of infinite-dimensional 
coupled equations since both the Floquet numbers, $p$ and $q$ 
may assume any integral values in between  
$-\infty$ and $+\infty$. But, if the frequency of the incident 
light is greater than that of corresponding 
band-width of the system, the Floquet sub-bands 
become decoupled. In this limit, one can apply 
Floquet-Magnus expansion \cite{Polkovnikov} to describe the 
system in terms of an effective Hamiltonian,  
 $H_{\rm eff}$, which can be written as
 $H_{\rm eff}=\sum_{i}\frac{H_{\rm eff}^i}{{\Omega}^i}$
where $H_{\rm eff}^0=H^0$ and $H_{\rm eff}^1=\sum_{p=1}^{\infty}\frac{1}{p}\left[H_{p},H_{-p}\right]$.
$H_{p} $ is the $p$-th order Fourier component.
Ignoring the higher order components ($|p|\geq 2 $) for high frequency,
$H_{\rm eff}$ is finally given by
\begin{equation}
\begin{aligned}
 H_{\rm eff} = H_{0}+\frac{1}{\Omega}\left[H_{1},H_{-1}\right]. 
 \label{6}
 \end{aligned}
\end{equation}
Throughout this investigation, frequency of the light is kept greater 
than that of the band-width of the undriven system, {\em i.e.}, 
the system is always kept 
in the off-resonant regime. Although
the electrons cannot be excited by direct absorption of this kind of irradiation, 
but instead the light is capable to modify the single-electron bands through
virtual photon-absorption processes \cite{Kitagawa2,Ezawa}. In Eq \ref{6}, 
$H_{0}$ describes the system 
where no photon exchange takes place, 
while $H_1$ ($H_{-1}$) takes into account the 
emission (absorption) of one virtual photon. 

After obtaining the expression of $ H_{0},H_{1}$ and $H_{-1}$, 
the effective time-independent Hamiltonian $H_{\rm eff}$ 
for the system is written as 
$ H_{\rm eff} = -\sum_{\textbf{k}} \psi_{\textbf{k}}^{\dagger} 
H_{\rm eff}(\textbf{k}) \psi_{\textbf{k}} $. 
The derivation of $H_{\rm eff}$ is available in Appendix \ref{10}. 
The diagonal elements of $H_{\rm eff}(\textbf{k})$ are zero. The 
off-diagonal upper-triangular elements of 
$H_{\rm eff}(\textbf{k})$ are written below. 
\begin{equation}
\begin{aligned}
 H_{\rm eff}^{12}(\textbf{k}) &= J_{0}(A_0)t_1 
-4iPt_1t_2+2iQt_1^{2}(e^{ik_1}+e^{ik_2}),\\
 H_{\rm eff}^{13}(\textbf{k}) &= J_{0}(\sqrt{2}A_0)t_2+J_{0}(A_0)t_1e^{ik_2},\\
 H_{\rm eff}^{14}(\textbf{k}) &= J_{0}(A_0)t_1 +4iPt_1t_2-2iQt_1^{2}(e^{-ik_1}+e^{ik_2}),\\
 H_{\rm eff}^{23}(\textbf{k}) &= J_{0}(A_0)t_1 -4iPt_1t_2+2iQt_1^{2}(e^{-ik_1}+e^{ik_2}),\\
 H_{\rm eff}^{24}(\textbf{k}) &= J_{0}(\sqrt{2}A_0)t_2+J_{0}(A_0)t_1e^{-ik_1},\\
 H_{\rm eff}^{34}(\textbf{k}) &= J_{0}(A_0)t_1 -4iPt_1t_2+2iQt_1^{2}(e^{-ik_1}+e^{-ik_2}).\nonumber
 \end{aligned}
\end{equation}
$H_{\rm eff}^{pq}(\textbf{k})$ are the elements of $p$-th row and $q$-th column of 
the $4\times4$ Hamiltonian, $H_{\rm eff}(\textbf{k})$,
where the constants $P $ and $Q$ are given by
\begin{equation}
\begin{aligned}
 P &= J_{1}(A_0)J_{1}(\sqrt{2}A_0)\sin(\pi/4)/\Omega, \\
 Q &= {J_{1}}^{2}(A_0)\sin(\pi/4)/\Omega. \\
 \end{aligned}
\end{equation}
Here $J_{n}(x)$ is the $n$-th order Bessel function given by 
$J_{n}(x) = \frac{1}{2\pi}\int_{0}^{2\pi}e^{-i(n\tau-x\sin(\tau)}d\tau $.

It is evident that the circularly polarized light breaks the TRS 
of the undriven system since  
$H_{\rm eff}(\textbf{k})\neq H_{\rm eff}^{\ast}(-\textbf{k})$. 
As soon as the periodic drive is turned on, {\em i.e.}, $A_{0}$ 
becomes non-zero, gaps appear in the band structure. 
At the same time, the system becomes topologically nontrivial in a sense that 
every band emerges with a definite value of Chern number. 
Band diagrams obtained by numerically diagonalizing the effective Hamiltonian
$H_{\rm eff}(\textbf{k})$ are shown in 
Fig \ref{band3d} (b), (c) and (d) for few particular sets of
parameter, $t_2,A_0$ and $\Omega$. 
The Chern numbers of respective 
bands are also noted. The sets of parameters are chosen in such 
a way that distribution of $C$ for the 
four energy bands are different. In those figures, `Energy' actually means 
the quasi-energies of the effective Hamiltonian since $A_{0}$ is non-zero. 
With the increase of $A_0$ beyond zero, both the flat-bands begin 
to become dispersive. 
At the same time, the remaining two bands get modified 
in such a fashion that a series of topological phase transitions occur, 
which will be described in detail in the 
next section. Dispersion relations along the path connecting the 
high-symmetry points of BZ are shown in Fig \ref{band2d} (b), (c) and (d). 

The system exhibits true and pseudo gap in the band diagram 
for different parameter regimes and the corresponding topological phases are noted as CI and
CSM\cite{Lieb}, respectively, depending on the lattice fillings.
True and pseudo band-gaps are defined in the following way.
If $E_{m}(\textbf{k}) \neq E_{n}(\textbf{k}^\prime)$ for all $m \neq n$ and $\textbf{k}$, $\textbf{k}^\prime$, then
true gap exists between $m$ and $n$-th energy bands. The gap is marked pseudo when
$E_{m}(\textbf{k}) \neq E_{n}(\textbf{k})$ for all $m \neq n$ and $\textbf{k}$. In the later case,
$E_{m}(\textbf{k}) = E_{n}(\textbf{k}^\prime)$ is a possible scenario. 
For both the cases, Chern numbesr are defined since the bands do not touch each other.  
The value of lattice filling determines the position of 
Fermi energy in the band diagram.  

However, Figs \ref{band3d}(d) and \ref{band2d}(d) indeed reveal the occurrence of
band inversion. 
This is due to the fact that dominant contribution to the band energy 
in the high-frequency regime is given by 
the term $H_0$. So, as $A_0$ crosses the value 2.4, the zeroth 
order Bessel function becomes negative, which is responsible 
for the inversion of the bands. 
The effect of linearly polarized light whose vector potential is given by
$\textbf {A} = A_0\left[\sin(\Omega t),0 \right] $ is examined 
on the system. But the system remains trivial, which is 
similar to the outcome of previous investigation \cite{Fiete1}. 
This is because of the fact that 
linearly polarized light fails to break the TRS of the system due to 
the combined effect of equal contribution and mutually compensating act  
of left and right circular states of polarization. 
However, the energy bands are found to get modified and gaps emerge in the spectrum
with the variation of both the parameters $A_0$ and $t_2$ 
in the presence of linearly polarized light. 

\section{Topological Properties}
\label{Properties}
\subsection{Chern numbers and Hall conductivity at zero temperature} 
The values of Chern numbers and Hall conductivity at zero temperature 
have been obtained in order to study the topological properties of the system
by considering the effective time-independent Hamiltonian in off-resonant regime.
Therefore, formalism developed for undriven system has been applied to 
obtain berry-curvature and zero-temperature Hall-conductivity, $\sigma_{H_{\rm eff}}(E)$ 
\cite{Owerre2}. $\sigma_{H_{\rm eff}}(E)$ actually corresponds to the 
dc component of optical Hall conductivity in the Floquet theory \cite{Mitra1,Fiete1}. 
Here $\sigma_{H_{\rm eff}}(E)$ is estimated numerically 
by using the Kubo formula \cite{TKNN},  
\begin{align*} 
\sigma_{H_{\rm eff}}(E) &= \!\frac{ie^2\hbar}{A}\sum_{\textbf{k}}\sum_{E_m<E<E_n}\!\!\!\!\!\!\!\!\nonumber\\
 & \frac{\langle m|v_1|n\rangle 
\langle n|v_2|m\rangle\!-\!\langle m|v_2|n\rangle 
\langle n|v_1|m\rangle}{(E_m-E_n)^2},
\end{align*}
where $\ket{l}=\ket{u_{l,\boldsymbol k}}, H_{\rm eff}(\boldsymbol k)\ket{l}=E_l\ket{l}$ and $l=m,n$. 
$A$ is the area of the system.  
$E$ is treated as the Fermi energy of the system. 
The velocity operator, $v_\alpha=(1/i\hbar)[\alpha,H_{\rm eff}]$ 
where $\alpha=1,2$ denote the $a_1$ and $a_2$ directions respectively. 
When $E$ falls in one of the band gaps, expression of 
$\sigma_{H_{\rm eff}}$ looks like,  
\be 
\sigma_{H_{\rm eff}}\left(E\right)=\frac{e^2}{h}\sum_{E_n<E}C_n. 
\label{hall-plateau}
\ee

\begin{figure*}[t]
\centering
\psfrag{E}{$E$}
\psfrag{sigma}{$\sigma_{H_{\rm eff}}$ ($e^2/h)$ and DOS}
\includegraphics[width=15cm]{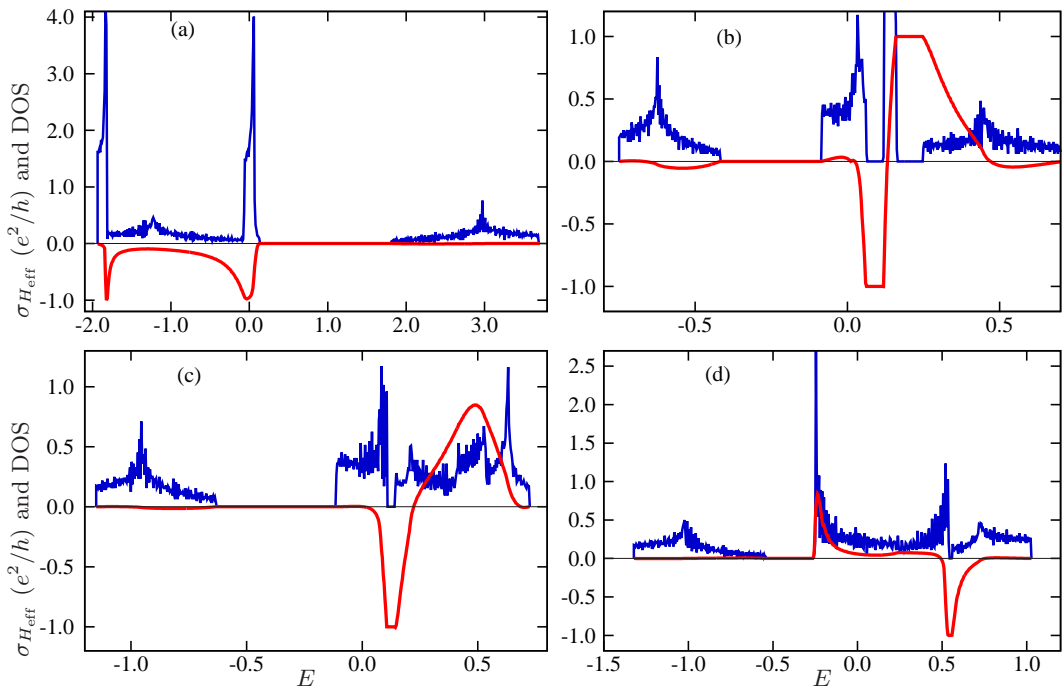}  %
\caption{(Color online) Topological phases in the presence of circularly polarized light 
with different $A_0$ when $t_2=1.0$, and $\Omega=10$. (a) $A_0=0.5$ (b) $A_0=2.1$
(c) $A_0=3.0$ (d) $A_0=3.6$. The Hall conductance $\sigma_{H_{\rm eff}}$ (red line) 
and DOS (blue line) are plotted
with respect to the Fermi energy, $E$.}
\label{hall}
\end{figure*}
Here, $C_n$ is the Chern number of $n$-th completely filled band, 
which is given by 
\begin{equation}
 \begin{aligned}
  C_n &=\frac{1}{2\pi}\int_{BZ}d^2\boldsymbol k \cdot \boldsymbol F_n\left(\boldsymbol k \right),
    \label{Cn}
 \end{aligned}
\end{equation}
where $\boldsymbol F_m(k)$ is the berry-curvature. The normal-component 
of  $\boldsymbol F_m\left(\boldsymbol k \right)$ is given by $F^\perp_m\left(\boldsymbol k \right)=
-i\left(\braket{\partial_{1} u_{m,\boldsymbol k}|\partial_{2} u_{m,\boldsymbol k}}-
    \braket{\partial_{2} u_{m,\boldsymbol k}|\partial_{1} u_{m,\boldsymbol k}}\right)$. 
Here $\ket{u_{m,\boldsymbol k}}$ are the eigenvectors of $H(\boldsymbol k)$ 
and $\partial_{j}=\frac{\partial}{\partial k_j}$.
In this numerical estimation, discretized version of Eq. \ref{Cn} introduced 
before by Fukui and others has been used \cite{Suzuki}. 
    
By keeping the frequency fixed at an off-resonant value, $\Omega=10$,
characteristics of topological phase transitions in the parameter space spanned
by $t_2$ and $A_0$ have been investigated.
Robustness of those results has been confirmed by fixing $\Omega$ to higher values. 
Now, the nature of phase transitions along two definite lines in the parameter-space 
will be described. First line is defined by fixing $t_2=1.0$. 
When $A_0$ is zero, Chern numbers are not defined for the lower three bands 
since there is no band gap.  Value of $C$ for the isolated upper band is found zero. 
For the line segment bounded by the limits $0.0<A_0\leq1.2$, 
the system remains in a non-trivial
topological phase with a definite value of $C$ for each band. 
Here, $C=-1,0,1,0$ for the bands arranged in 
the ascending order of energy. The system mostly remains in
CSM phase for 1/2 and 1/4 lattice fillings. 

From Fig \ref{hall}(a), it is observed that there are three plateaus for 
$\sigma_{H_{\rm eff}}=n(e^2/h)$ with $n=-1,-1,0$. Plateaus are not always 
prominent due to 
the presence of either pseudo or narrow gaps in the band diagram. 
The density of states (DOS) exhibits sharp peaks around the 
energies where $\sigma_{H_{\rm eff}}(E)$ undergoes sudden rise and fall.
 
It is worth mentioning that for infinitesimally small values of $A_0$, 
the original pair of flat bands develop very low curvature such that 
corresponding band-widths are vanishingly small. 
At this limit, they acquire non-zero Chern numbers, $C=\pm 1$,  but at the 
same time the band-gaps are too narrow to have a considerable value of 
flatness ratio.  

With further increase of $A_0$ and when $1.2<A_0\leq2.0$, 
a series of topological phase transitions is observed due to the 
rapid occurrence of intermediate 
band-touchings and gap-openings with the change of amplitude. 
A closer look reveals that 
gaps close and reopen when $A_0$ crosses 1.2  
along with the emergence of a new topological phase with 
$C=1,-2,1,0$. This is a quadratic band-touching point between the two lower 
bands as the Chern numbers of these bands are exchanged 
by $\pm 2$ \cite{Kundu}. Upon further increase of $A_0$, 
a number of topological phases appear where the Chern numbers are 
redistributed consecutively as $(1,-2,0,1)$, $(-1,0,0,1)$, 
$(-1,-1,1,1)$, $(0,-2,1,1)$, $ (0,-2,3,-1)$ and $(0,-1,2,-1)$.  
The intermediate phase transitions involve with the 
emergence of either Dirac or quadratic band touching points. 

At $A_0=2.1$, the system again exhibits true band-gaps where the Chern numbers of 
the bands are $(0,-1,2,-1)$. The corresponding figure 
(Fig \ref{hall}(b)) shows that there are three prominent plateaus 
in $\sigma_{H_{\rm eff}}=n(e^2/h)$ with $n=0,-1,1$. The system remains in CI 
phase for 1/2 and 1/4 lattice fillings. 
When $A_0$ is increased to 2.12, a new topological phase with $C=0,1,1,-2$ 
appears. 
At the subsequent transition point, the gap between middle two bands closes. 
Dispersion relations of the two bands are linear at the touching point. with slight increase of $A_0$, that gap reopens and the 
Chern numbers of those two middle bands are exchanged by $\pm 1$ 
leading to a new topological phase with $C=0,0,2,-2$. 

At the points $A_0=2.4$ and $2.5$, the system 
drives into the trivial phase since all the Chern numbers become zero. 
This happens because of the fact that 
around those points,
the value of zeroth order Bessel function becomes vanishingly small. 
When $A_0$ reaches to the value 2.6, the Chern numbers are restored to $(0,0,2,-2)$.
The system again undergoes a series of rapid phase transitions with further modulation of $A_0$. 
Two different phases with $C=0,-1,2,-1$ at
$A_0=3.0$ and $C=0,1,-2,1$ at $A_0=3.6$ are shown in Fig \ref{hall} 
(c) and (d), respectively. For $A_0=3.0$, the plateau at $n=-1$ is prominent
as the system lies in CI phase for 1/2 lattice filling. 
Elsewhere, the system remains in CSM phase. 

Another series of phase transitions can be obtained if the parameter space 
is explored along a different line by fixing $A_0=2.1$.  
When $0\leq t_2\leq 0.2$, the system is in Chern insulating phase 
for all values of lattice fillings with $C=-1,-1,1,1$. Then the gap
between the lower two bands closes and reopens to 
redistribute the Chern numbers as $(0,-2,1,1)$ at $t_2=0.3$. 
For $0.4<t_2<0.5$, a phase $C=0,-2,3,-1$ appears due to quadratic
closing of lower two bands.
Then, we obtain fully gapped CI phase with $(0,-1,2,-1)$
for $0.5\leq t_2\leq 1.1$.
Thereafter, the system undergoes rapid transitions
through the phases $(0,1,0,-1),(0,1,1,-2),(0,0,2,-2)$. Those
can be identified as CSM phase because the system 
demonstrates several pseudo-gaps in its band diagram.
For $t_2>1.5$, system becomes trivial as all the Chern numbers become zero. 

\begin{figure}[t]
\centering
\includegraphics[width=8cm,height=5cm,trim={0.0cm .5cm 0.0cm 0.0cm}]{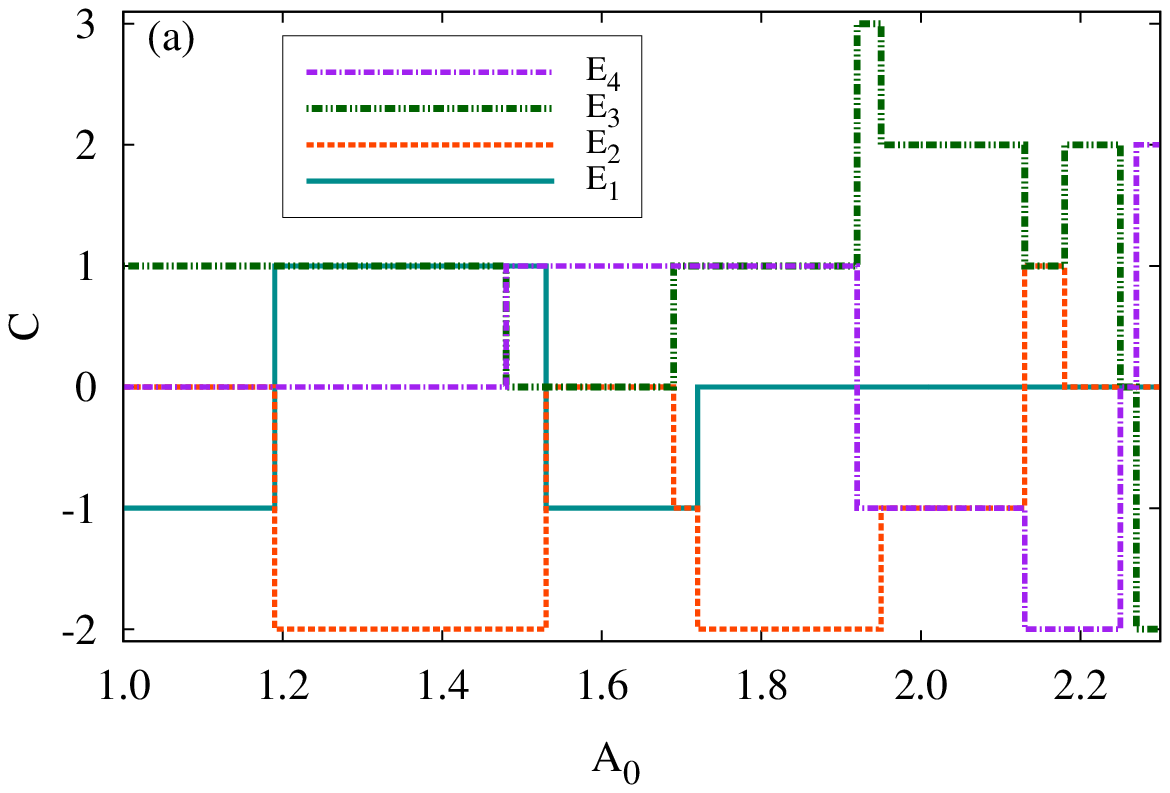}  %
\includegraphics[width=8cm,height=5cm,trim={0.0cm .5cm 0.0cm 0.0cm}]{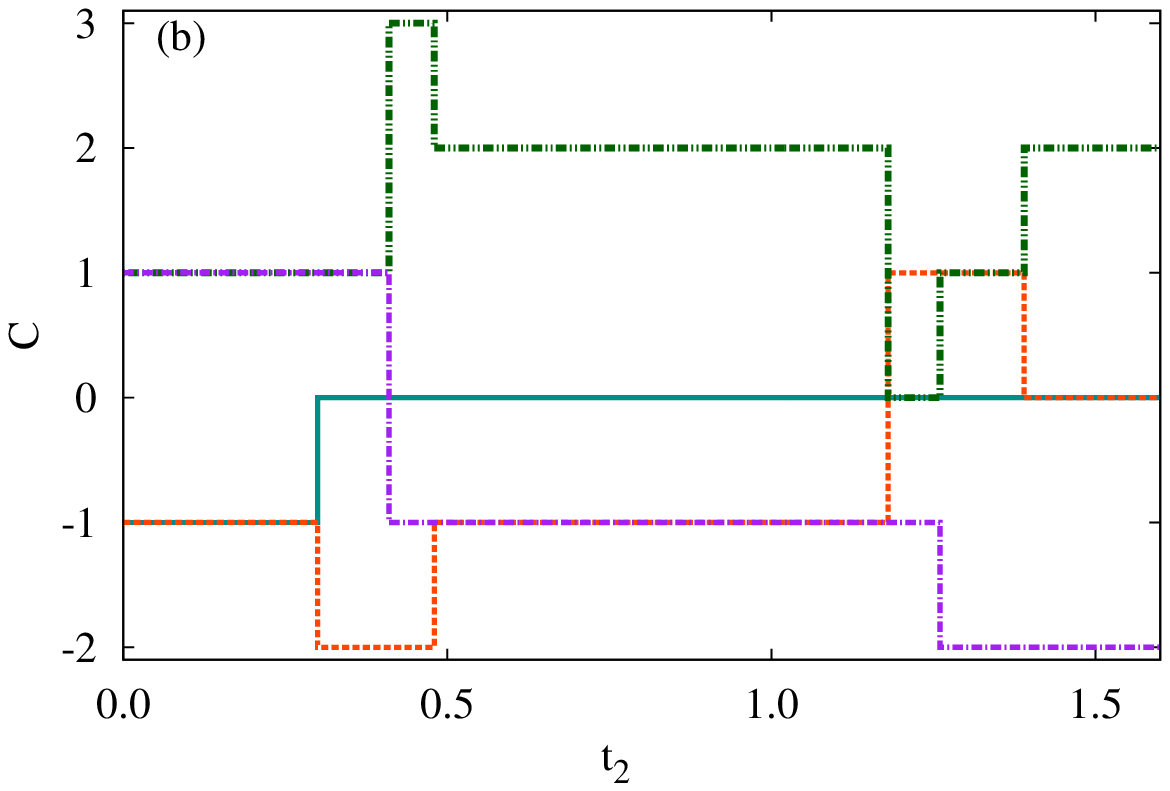} 
\caption{(Color online) (a) Topological phase diagram of the system 
shows rapid variation of phases with $A_0$, where  
$t_2$ and $\Omega$ are fixed at 1.0 and 10.0, respectively. 
Chern numbers of the four bands are plotted as a function of $A_0$.
(b) Topological phase diagram of the system with respect to $t_2$. $A_0$ and $\Omega$ are 
fixed at 2.1 and 10.0, respectively. Summation of the Chern numbers over all the bands are 
zero, which is evident from the figures.
}
\label{phase}
\end{figure}

Variation of topological phases as well as distribution of $C$ for respective 
phases with the change of $A_0$ and $t_2$ are shown in Figs \ref{phase} (a) 
and (b), respectively. 
Rate of variation of phases with respect to $A_0$ is faster. 
Maximum value of $C$ is $+3$ while minimum value of that is $-2$.    
Extent of phases is uneven and appearance of them is quite random. 
Multiple topological phases and rapid variations of them are the two 
special findings in this model. 
 
\subsection{Topological edge states}
\begin{figure}[]
\centering
\psfrag{x}{\text{\scriptsize{Right edge}}}
\psfrag{y}{\text{\scriptsize{Left edge}}}
\psfrag{q2}{\text{\scriptsize{$k_1$}}}
\psfrag{wave-square}{\text{ \scriptsize {$| \psi (k_1\!\!=\!\!0)|^2$}}}
\psfrag{Energy}{\text{\scriptsize{Energy($k_1$)}}}
\psfrag{site}{\text{\scriptsize{Site}}}
\psfrag{DOS}{\text{\tiny{DOS}}}
\includegraphics[width=8cm,height=7.0cm]{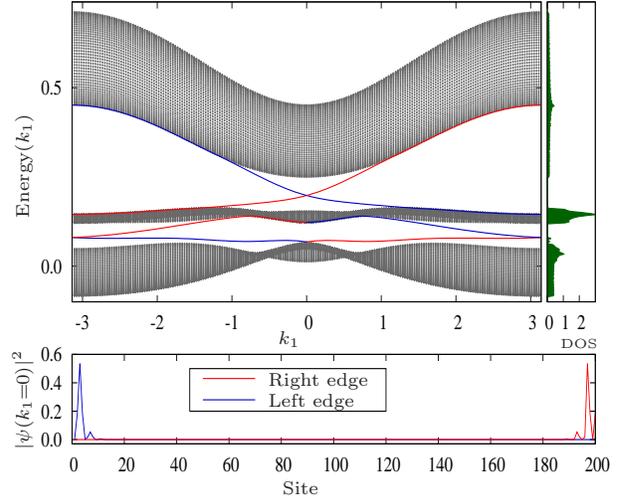} 
\caption{(Color online) Edge states of square-octagon lattice in the 
presence of circularly polarized light for the 
parameters $t_2=1.0,\Omega=10,A_0=2.1$ for $N=50$ unit cell along $k_2$.
Chern numbers of the upper three bands are $C={-1,2,-1}$. 
The lowest band with $C=0$ is not shown in figure since 
the lowest band-gap 
does not accommodate any edge states. The side-panel 
indicates the density of states. Lower panel shows 
the probability density of left (blue) and right 
(red) going edge states with respect to site number of 
the strip for $k_1=0$. The colour 
of the edge state energies corresponds to the colour of 
particular edge where the probability density is concentrated. }
\label{edge}
\end{figure}

In order to examine whether the system obeys the 
`bulk-boundary correspondence' rule, the effective Hamiltonian of 
a finite strip of the system is considered. 
Values of the parameter are taken as $t_2=1.0, A_0=2.1, \Omega=10$. 
The system remains in a particular CI phase for these values. 
To create edges in the system, periodic boundary condition along 
$\boldsymbol a_2$ direction is withdrawn, 
so that $k_2$ is no longer a good quantum number. Now, 
considering $N=50$ unit cells {\em i.e.} 200 
sites along $a_2$ direction, the resulting $4N \times 4N$ Hamiltonian 
is diagonalized. The quasi-energy eigenvalues of the effective Hamiltonian 
are plotted as a function of $k_1$ in Fig \ref{edge}. 
Density of states are also drawn in the side-panel
to show that the spectrum has a true gap.

Evidently, additional states are found to appear within the band gaps, 
those actually connect the distinct bulk bands. These are known as the 
edge states, which are indeed localized in either left (blue curves) 
or right (red curves) edge of the finite lattice, 
as shown in the lower panel of Fig \ref{edge}. 
It may be noted that the edge states are chiral in nature since 
the right-going states or the states
with positive group velocity are localized in the 
right edge, and similarly, the left-going states with negative
group velocity are always localized in the left edge.

Hence, the results are in accordance with the `bulk boundary correspondence' 
rule which states that:
sum of the Chern numbers up to the $i$-th band, 
$\nu_i = \sum_{j\leq i}C_j $ is equal to the 
number of pair of edge states in the gap \cite{Mook}. 
It can be verified from the fact that 
the Chern numbers of the corresponding bands are 
${-1,2,-1}$ in the ascending order of energy as shown in Fig \ref{edge}. 
So, there must be
one pair of edge states in each band gap.

\section{Summary and Discussion}
\label{discussions}
The effect of circularly polarized light on the topology and band-structure
of square-octagon lattice is studied by using the Floquet-Bloch 
theory. The system is topologically trivial in 
the absence of light with a pair of flat bands and
quadratic band-touching points in its band-diagram. As soon as the 
system is exposed to circularly polarized light, flat bands become dispersive
and the system is driven into 
non-trivial topological phase at the same time. 
Photo induced band structure exhibits true and pseudo gaps 
depending on the amplitude of light which eventually leads to the 
emergence of multiple topological phases with a variety of 
Chern number distribution. 
Thus a series of topological 
phase transitions is observed with the modulation of 
amplitude of the incident light and hopping strengths. 
The overall state of 
the system is identified either CSM or 
CI phase depending on the value of lattice filling. 
Those phases are topologically 
robust in a sense that no further change
is observed with the variation of frequency as long as 
it is fixed in off-resonant regime. The frequency is always kept 
higher than the band-width of the undriven system to make sure that Floquet-Magnus 
expansion is always valid. 
Topological properties are characterized in terms of Chern numbers of 
distinct energy bands which is further justified with the behaviour of 
Hall conductance and evidence of edge states. 
Linearly polarized light has no effect on 
the topological properties of the system as it 
fails to break the TRS.

Previous investigation reveals that Dirac cones appear at the 
$\Gamma$ and M points in the absence of NNN hopping \cite{Ueda}. 
The present study indicates that the system 
exhibits photo-induced CI phase in this case, 
for any values of lattice filling. 
So, it would be interesting to find the photo-induced topological 
properties of a particular model with different values of 
hopping strengths for intra- and inter-square-plaquette NN bonds 
on the square-octagon lattice. 
In this context, 
computation of finite-frequency optical conductivity will become 
useful for the experimental detection of topological phases.
Additionally, further expansion of 
this work can be made by invoking further neighbour hoppings 
in this tight-binding model. 

FTI phase has been observed on graphene 
with the help of opto-helical wave-guides \cite{Rechtsman} 
and Floquet-Bloch bands have been found on the 
surface of the compound, Bi$_{2}$Se$_3$ \cite{Wang2}. 
In another development, optical lattices with ultracold atoms 
for two-dimensional honeycomb \cite{Panahi}, 
checkerboard \cite{Wirth} and kagom\'e \cite{Jo} structures 
have been realized. So, the square-octagon optical lattice 
will hopefully  become realizable in near future which, 
on the other hand, will pave the way for observation of these topological phases. 
However, square-octagon lattice based on real material had been come to light 
long ago in the context of antiferromagnetic compound, CaV$_4$O$_9$ \cite{Taniguchi}. 
The spin-1/2 V$^{4+}$ ions in this spin-liquid 
constitute  square-octagon lattice structure. 
Meanwhile, quasi 
square-octagon structure has been found in $(10\bar{1}0)$ surface of
functional material ZnO \cite{He}.
So, the recent trends of investigation indicate the realization of these findings very soon. 
\section{ACKNOWLEDGMENTS}
AS acknowledges the CSIR fellowship, no. 09/096(0934) (2018), India.  
AKG acknowledges BRNS-sanctioned 
research project, no. 37(3)/14/16/2015, India.
  
\appendix
\section{Derivation of $H_{\rm eff}$}
 \label{10}
 To obtain the Fourier components of $H({\textbf{k}},t)$ (Eq \ref{Ht}),
 the coefficient of a particular term 
$-t_1 \sum_{\textbf{k}}A_{\textbf{k}}^{\dagger}B_{\textbf{k}}$ in the expression of 
 $H_n({\textbf{k}})$ becomes 
   \bea
      &&\frac{1}{T}\int_{0}^{T}dt e^{-in\Omega t}e^{-\frac{iA_0}{\sqrt{2}}(\sin(\Omega t)+\cos(\Omega t)}\nonumber\\
     && = \frac{1}{T}\int_{0}^{T}dt e^{-in\Omega t}e^{-iA_0\sin(\Omega t +\pi /4)}, \nonumber\\
     && = \frac{1}{2\pi}\int_{0}^{2\pi}d\tau e^{-in\tau}e^{in\pi/4}e^{-iA_0\sin(\tau)}, \nonumber\\
     && = J_n(-A_0) e^{in\pi/4},
   \eea
where $\tau=\Omega t +\pi/4$ and $\Omega T=2\pi$.
Similarly, the coefficient of $-t_1 \sum_{\textbf{k}}B_{\textbf{k}}^{\dagger}A_{\textbf{k}}$ becomes $J_n(A_0) e^{in\pi/4}$.
In the same way, all the other coefficients can be derived. The matrix elements of 
$4\times 4$ matrix $H_{n}(\textbf{k})$ are explicitly given by
 \begin{equation}
\begin{aligned}
 H_{n}^{12}(\textbf{k}) &= t_1 J_n(-A_0) e^{in\pi/4},\\
 H_{n}^{21}(\textbf{k}) &= t_1 J_n(A_0) e^{in\pi/4},\\
 H_{n}^{13}(\textbf{k}) &= t_2 J_{n}(-\sqrt{2}A_0) e^{in\pi/2}+J_{n}(A_0)t_1e^{i(n\pi/2+k_2},\\
 H_{n}^{31}(\textbf{k}) &= t_2 J_{n}(\sqrt{2}A_0) e^{in\pi/2}+J_{n}(-A_0)t_1e^{i(n\pi/2-k_2},\\
 H_{n}^{14}(\textbf{k}) &= t_1 J_{n}(A_0) e^{-in\pi/4},\\
 H_{n}^{41}(\textbf{k}) &= t_1 J_{n}(-A_0) e^{-in\pi/4},\\
 H_{n}^{23}(\textbf{k}) &= t_1 J_{n}(A_0) e^{-in\pi/4},\\
 H_{n}^{32}(\textbf{k}) &= t_1 J_{n}(-A_0) e^{-in\pi/4},\\
 H_{n}^{24}(\textbf{k}) &= t_2 J_{n}(\sqrt{2}A_0)+t_1 J_{n}(-A_0)e^{-ik_1},\\
 H_{n}^{42}(\textbf{k}) &= t_2 J_{n}(-\sqrt{2}A_0)+t_1 J_{n}(A_0)e^{ik_1},\\
 H_{n}^{34}(\textbf{k}) &= t_1 J_{n}(A_0) e^{in\pi/4},\\
 H_{n}^{43}(\textbf{k}) &= t_1 J_{n}(-A_0) e^{in\pi/4},\\
 \label{8}
 \end{aligned}
\end{equation}
 Now, applying the relations $J_{-n}(A_0)={(-1)}^{n}J_{n}(A_0)$ and 
$J_{n}(-A_0)={(-1)}^{n}J_{n}(A_0)$, matrix elements for 
the Fourier components of $ H_{0}(\textbf{k})$, $H_{1}(\textbf{k})$ and $H_{-1}(\textbf{k})$ 
have been obtained and finally, $H_{\rm eff}$ is obtained by Eq \ref{6}.

\end{document}